# Synaptic effects on the intermittent synchronization of gamma rhythms

Quynh-Anh Nguyen[1*†], Leonid L Rubchinsky[1,2‡]

[1]Department of Mathematical Sciences, Indiana University Purdue University Indianapolis, Indianapolis, IN 46202, USA
[2]Stark Neurosciences Research Institute, Indiana University School of Medicine, Indianapolis, IN 46202, USA

† ORCID: 0000-0003-2843-1841
‡ ORCID: 0000-0002-6596-4594

*Current address: Department of Mathematical Sciences, University of Indianapolis, Indianapolis, IN 46227, USA

**\* Correspondence:**
Corresponding Author
nguyenapp@uindy.edu

**Abstract**

Synchronization of neural activity in the gamma frequency band is associated with various cognitive phenomena. Abnormalities of gamma synchronization may underlie symptoms of several neurological and psychiatric disorders such as schizophrenia and autism spectrum disorder. Properties of neural oscillations in the gamma band depend critically on the synaptic properties of the underlying circuits. This study explores how synaptic properties in pyramidal-interneuronal circuits affect not only the average synchronization strength but also the fine temporal patterning of neural synchrony. If two signals show only moderate synchrony strength, it may be possible to consider these dynamics as alternating between synchronized and desynchronized states. We use a model of connected circuits that produces pyramidal-interneuronal gamma (PING) oscillations to explore the temporal patterning of synchronized and desynchronized intervals. Changes in synaptic strength may alter the temporal patterning of synchronized dynamics (even if the average synchrony strength is not changed). Larger values of local synaptic connections promote longer desynchronization durations, while larger values of long-range synaptic connections promote shorter desynchronization durations. Furthermore, we show that circuits with different temporal patterning of synchronization may have different sensitivity to synaptic input. Thus, the alterations of synaptic strength may mediate physiological properties of neural circuits not only through change in the average synchrony level of gamma oscillations, but also through change in how synchrony is patterned in time over very short time scales.

## 1. INTRODUCTION

Synchronization is a widespread phenomenon in neuroscience, and it likely plays a crucial role in brain functions through collectively established neuronal behavior (e.g., Buzsaki 2006). Synchronization rarely stays perfect over prolonged intervals of time. Moreover, imperfect and transient synchronization may be functionally effective (e.g., Borgers et al., 2014; Palmigiano et al., 2017) and there are several considerations of how imperfectly synchronized regimes may arise in neural systems (e.g., Chouzouris et al., 208; Bansal et al., 2019; Kanagaraj et al., 2024). If synchrony is not perfect, there are intervals of time when signals are in (or very close to) the synchronized state, and intervals of time when signals are not close to the synchronized state. Even though synchronization of oscillations is not an instantaneous phenomenon as it implies repetitive temporal coordination of oscillatory signals, it is possible to consider if signals are synchronized or not at each cycle of oscillations (as long as there is a statistically significant synchrony overall). Data analysis techniques to analyze this temporal patterning of synchronization on very short time scales have been developed over the last decade (e.g., Park et al., 2010; Ahn et. al., 2011; Ahn and Rubchinsky 2013). This analysis of synchronization traces how signals get in and out of synchronized state in time. It allows ones to distinguish cases of many short desynchronizations, a few long desynchronizations, and a wide spectrum of possibilities in between them, even if they all have the same average synchronization strength. The latter situation – same average synchronization, but different temporal patterning of synchronized dynamics – has been observed in several experiments (Ahn et. al., 2014, 2018).

Neural synchronization in the gamma frequency band has been studied extensively because of its significance for facilitation of interneuronal communication, memory formation, and a variety of cognitive phenomena (e.g., Fell and Axmacher, 2011; Colgin, 2011; Buzsaki and Schomburg, 2015; Fries, 2015). Abnormalities of gamma rhythm synchronization are related to schizophrenia (e.g., Uhlhaas and Singer, 2010, 2015; Spellman and Gordon, 2015; Pittman-Polletta et al., 2015) as well as to autism spectrum disorders (e.g., Sun et al., 2012; Malaia et al., 2020). Thus, the objective of this study is to use computational neuroscience techniques to look at the gamma rhythm synchronization in the context of imperfect synchronization and the temporal patterning of synchronous activity.

Properties of gamma-band oscillations in cortical circuits substantially rely on the properties of cortical synapses (e.g., Ermentrout and Kopell, 1998; Buzsáki and Wang, 2012; Salkoff et. al., 2015; Borgers, 2017). Thus, we consider how synaptic changes may affect the temporal patterning of synchronized activity. We approach the study of synchronization dynamics from time-series analysis angle (which may make it more experimentally relatable). We use a medium-size conductance-based pyramidal-interneuron gamma (PING) network to investigate the effects of synaptic connections on the temporal dynamics of synchronization. A preliminary study with a very small network of excitatory and inhibitory neurons suggested that variation of synaptic strength does affect the temporal patterning of imperfect synchronization (Nguyen and Rubchinsky, 2021). Moreover, there were suggestions that different temporal patterning can affect how a partially synchronized network responds to an external input even if the average synchronization strength is the same (Ahn and Rubchinsky, 2017), so we consider this issue here too. Given the importance of gamma-band synchronization for many neural functions, understanding the mechanisms and effects of different temporal patterning of this synchronization will advance comprehension of the connections between neural physiology and neural functions.

## 2. METHODS

### 2.1 Model neurons

The network has two circuits with models following (Borgers et. al., 2012; Borgers, 2017). While there are different modeling approaches to the synchronized gamma rhythm, this modeling choice provides a reasonable combination of simplicity and complexity as it retains the kinetics of membrane currents facilitating gamma activity. Each circuit has a population of pyramidal neurons and a population of interneurons, see Fig. 1A. Each neuron is modeled by a single compartmental conductance-based model. The equation assumes that the cell membrane behaves like a capacitor ($CV = Q$ or $C\frac{dV}{dt} = I_{membrane}$). Thus, the transmembrane voltage has the below formulation:

$$C_m \frac{dV}{dt} = -I_{Na} - I_K - I_L - I_{syn} + I_{app} \quad (1)$$

The transient sodium current is given by $I_{Na} = g_{Na}^*(V - v_{Na}) = g_{Na}m^3h(V - v_{Na})$ where $g_{Na}^*$ is the (voltage-dependent) conductance of the sodium channel, $g_{Na}$ is maximal conductance, and $v_{Na}$ is the reversal potential, $m$ and $h$ are voltage-dependent gating variables.

The timescale of the activation variable $m$ is significantly smaller than the time scale of other activation variables; $m$ is assumed instantaneous and thus substituted by its steady-state function, $m = \alpha_m(V)/(\alpha_m(V) + \beta_m(V))$, with

$$\alpha_m(V) = \frac{0.32(V+54)}{1-\exp\left(-\frac{V+54}{4}\right)} \quad , \quad \beta_m(V) = \frac{0.28(V+27)}{\exp\left(\frac{V+27}{5}\right)-1} \quad (2)$$

for excitatory neurons (Traub and Miles, 1991) and

$$\alpha_m(V) = \frac{0.1(V + 35)}{1 - \exp\left(-\frac{V + 35}{10}\right)} \quad , \quad \beta_m(V) = 4\exp\left(-\frac{V + 60}{18}\right) \tag{3}$$

for inhibitory neurons (Wang and Buzsáki, 1996).

The inactivation variable $h$ obeys the first-order kinetics:

$$\frac{dh}{dt} = \alpha_h(V)(1 - h) - \beta_h(V)h \tag{4}$$

with

$$\alpha_h(V) = 0.128\exp\left(-\frac{V + 50}{18}\right) \quad , \quad \beta_h(V) = \frac{4}{1 + \exp\left(-\frac{V + 27}{5}\right)} \tag{5}$$

for excitatory neurons (Traub and Miles, 1991) and

$$\alpha_h(V) = 0.35\exp\left(-\frac{V + 58}{20}\right) \quad , \quad \beta_h(V) = \frac{5}{1 + \exp\left(-\frac{V + 28}{10}\right)} \tag{6}$$

for inhibitory neurons (Wang and Buzsáki, 1996). $\alpha_h$ can be interpreted as the time rate such that the closed gates open, and $\beta_h$ is the time rate such that the open gates close.

The persistent potassium current is described as $I_K = g_K^*(V - v_K) = g_K n^4 (V - v_K)$ where $g_K^*$ is the potassium channel conductance, $g_K$ is maximal conductance, and $v_K$ is the reversal potential.

The activation function $n$ obeys first-order kinetics:

$$\frac{dn}{dt} = \alpha_n(V)(1 - n) - \beta_n(V)h \tag{7}$$

with

$$\alpha_n(V) = \frac{0.032(V + 52)}{1 - \exp\left(-\frac{V + 52}{5}\right)} \quad , \quad \beta_n(V) = 0.5\exp\left(-\frac{V + 57}{40}\right) \tag{8}$$

for excitatory neurons (Traub and Miles, 1991) and

$$\alpha_n(V) = \frac{0.05(V + 34)}{1 - \exp\left(-\frac{V + 34}{10}\right)} \quad , \quad \beta_n(V) = 0.625\exp\left(-\frac{V + 44}{80}\right) \tag{9}$$

for inhibitory neurons (Wang and Buzsáki, 1996). Similar as above, $\alpha_n$ can be interpreted as the time rate such that the closed gates open, and $\beta_n$ is the time rate such that the open gates close.

Note that the formulation of $\alpha_h, \alpha_n, \beta_h,$ and $\beta_n$ in Equations (6) and (9) for inhibitory neurons mimics the fast-firing parvalbumin-positive (PV+) basket cells that play an important role in generating gamma rhythm because they can sustain high-frequency firing (Wang and Buzsáki, 1996).

Furthermore, $I_L = g_L(V - v_L)$ is the leak current where $g_L$ is the leak conductance and $v_L$ is the reversal potential. Next, $I_{app}$ is the injected current. Excitatory neurons have $C_m = 1\ \mu F/cm^2, v_{Na} = 50\ mV, v_K = -100 mV, v_L = -67 mV, g_{Na} = 100\ mS/cm^2, g_K = 80\ mS/cm^2, and\ g_L = 0.1\ mS/cm^2$ (Traub and Miles, 1991).

Inhibitory neurons have $C_m = 1\ \mu F/cm^2, v_{Na} = 55\ mV, v_K = -90\ mV, v_L = -65\ mV, g_{Na} = 35\ mS/cm^2, g_K = 9\ mS/cm^2$, and $g_L = 0.1\ mS/cm^2$ (Wang and Buzsáki, 1996).

## 2.2 Model synapses

The synaptic current is $I_{syn} = g^*_{syn}(t)(V_{post} - v_{syn}) = g_{syn}s(t)(V_{post} - v_{syn})$. Here, $g^*_{syn}(t)$ is a voltage-dependent synaptic conductance, $V_{post}$ is the potential of the postsynaptic cell, and $v_{syn}$ is the reversal potential of the synapse. We set $v_{syn} = 0\ mV$ for the AMPA-receptor-mediated excitatory synapse and $v_{syn} = -80\ mV$ for the GABA-receptor-mediated inhibitory synapse (Borgers et al., 2012; Borgers, 2017).

$$\frac{ds}{dt} = H(V_{pre})\frac{1-s}{\tau_r} - \frac{s}{\tau_d} \tag{10}$$

Here, $H(V_{pre}) = (1 + tanh(V_{pre}/4))/2$ is instantaneous and of sigmoidal form with $H(V_{pre}) \approx 0$ when $V_{pre} \ll 0$ and $H(V_{pre}) \approx 1$ when $V_{pre} \gg 0$. The synaptic rising time is $\tau_r = 0.1\ ms$ for the AMPA-receptor-mediated excitatory synapse and $\tau_r = 0.3\ ms$ for the GABA-receptor-mediated inhibitory synapse. The synaptic decay time $\tau_d$ is $\tau_d = 3\ ms$ for the AMPA-receptor-mediated excitatory synapse and is $\tau_d = 9\ ms$ for GABA-receptor-mediated inhibitory synapse. These parameters are taken from (Borgers et al., 2012; Borgers, 2017).

## 2.3 Model network

The model consists of two interconnected networks. Each network has $N_E = 40$ excitatory neurons and $N_I = 10$ inhibitory neurons. The ratio of excitatory neurons over inhibitory neurons (40/10=4) is the approximate ratio of glutamatergic to GABAergic neurons in the cortex (Sahara et al., 2012). While a network of excitatory and inhibitory neurons can generate oscillations of various frequency, the firing frequency depends significantly on the strength of the inhibitory synaptic and external inputs. We describe how we chose the values of synaptic strength and external drive in the next three paragraphs.

The synaptic strength is $g_*$ for a connection between neurons within a circuit and $c_*$ for a connection between neurons from different circuits. The presence of coupling between each neuron is randomly assigned with the probability $P_*$ for within-network connection and $p_*$ for cross-network connection, see Fig. 1A. For each combination of parameters, we generated 50 random connection matrices, simulated these 50 networks, and analyzed the mean results. To make sure that local connections are more dense than distant connections, we set $P_* = 4p_*$. The symbol * stands for an excitatory-to-inhibitory connection (EI), an inhibitory-to-excitatory connection (IE), and an inhibitory-to-inhibitory connection (II). Thus, we have $(g_{EI}, g_{IE}, g_{II})$ and $(P_{EI}, P_{IE}, P_{II})$ as the connection strength and the connection probability, respectively, for intra-network coupling. Similarly, $(c_{EI}, c_{IE}, c_{II})$ and $(p_{EI}, p_{IE}, p_{II})$ are the connection strength and the connection probability, respectively, for inter-network coupling. The excitatory to excitatory connection does not play a significant role in gamma oscillation (Borgers, 2017; Ermentrout and Kopell, 1998); thus, we do not consider them in our model, neither within the networks nor between networks (so that the question of whether this coupling will affect the dynamics is left out of the present study, but the modeling approaches all synapses in a uniform manner). Additionally, there are no recurrent connections between the same neurons.

The default values for synaptic strength between neurons are $g_{EI} = 0.0125$, $g_{II} = 0.075$, $g_{IE} = 0.35$; $c_{EI} = 0.01$, $c_{II} = 0.04$, $c_{IE} = 0.04$. Connection probabilities are $P_{EI} = 0.4, P_{II} = 0.4, P_{IE} = 0.4$; $p_{EI} = 0.1, p_{II} = 0.1, p_{IE} = 0.1$. The schematic of the model is portrayed in Fig. 1A. A similar model structure has been used in (Borgers et. al., 2012; Borgers, 2017; Quax et al., 2017; Dumont and Gutkin, 2019).

Each neuron receives a constant current $I_{app}$ of different magnitude. It is drawn from Gaussian distribution:

$$I_{app} \sim \mu(1 + \mathcal{N}(0, \sigma)) \tag{11}$$

where $\mu$ and $\sigma$ are chosen differently for each population so that each network fires at a slightly different frequency. In the slower network, $\mu_E = 3.5, \sigma_E = 0.15$; $\mu_I = 0.25, \sigma_I = 0.2$. In the faster network, $\mu_E = 3.5, \sigma_E = 0.15$; $\mu_I = 0.2, \sigma_I = 0.2$.

The simulations are performed using adaptive Runge-Kutta-Fehlberg 45 (rkf45) method in the GNU Scientific Library of the software Brian 2 (Stimberg et. al., 2019) The network activity is recorded with time step of $\Delta t = 0.05\ ms$. An illustration of network activity from a particular simulation for the default parameter set is presented in Fig. 1B.

**2.4 Time series analysis**

For each network, we compute a mean-field – like quantity, which can be considered as a proxy for a local field potential (LFP), a signal which is commonly recorded in in vivo neurophysiological studies. While the actual origin of LFP may be a complicated matter, we will refer to this mean-field signal as LFP for the sake of simplicity. It is computed as the mean of the total synaptic current into a neuron:

$$LFP = \frac{1}{N}\sum_{i=1}^{N} I_{syn}^{i} \qquad (12)$$

where $N = N_E + N_I$, and $I_{syn}^i$ is the total synaptic current into neuron $i$. The mechanisms behind formation of LFP are complicated and may include factors beyond synaptic currents (e.g., Buzsaki et al., 2012). However, since our model is concerned with the synchronization between networks and does not have geometrically realistic neuroanatomy, this mean-field – like signal is selected as a network activity variable and is referred to in this paper as LFP. Figure 1C (gray line) illustrates an example of simulated LFP. This is not the only possible variable to use to study synchronization dynamics, but it lets one put this study in the context of experimental studies of the temporal patterning of synchrony as those mostly use mean field-like variables.

The LFP signals are band-pass filtered with a 20-60 Hz spectral window to detect the activity in the low gamma band. We use a Kaiser window digital finite impulse response (FIR) filter sampled at 10kHz; band-pass 20-60 Hz with ripple 5%, stop-band 0-2 Hz with ripple 1%. Then, zero-phase filtering is used to avoid phase distortion. Hilbert transform is implemented to compute the phase of the filtered signals; we call $\varphi_1(t)$ and $\varphi_2(t)$ the phase of LFP signals from circuits 1 and 2, respectively. The phase difference between the two signals at any time point $t$ can be represented as a vector $\xi(t) = e^{i(\varphi_1(t)-\varphi_2(t))}$ in the complex plane. Averaging these vectors across all time points yields the average synchronization index (or the strength of phase locking between the two signals) (Pikovsky et al., 2004; Hurtado et al., 2004), shown below

$$\gamma = \left\| \frac{1}{N_T} \sum_{t_j=1}^{N_T} \xi(t_j) \right\|. \qquad (13)$$

Here, $N_T$ is the number of timepoints, and $\|.\|$ is the modulus of a complex number. Figure 1C shows an example of sine of phase $\varphi_*(t)$ (in gray). The value of $\gamma$ falls in between 0 (no synchrony or complete lack of phase locking) and 1 (full synchrony or perfect phase locking). If the vectors $\zeta(t)$ point to similar direction (i.e. the phase differences $\varphi_1(t) - \varphi_2(t)$ are similar) for many timepoints $t$, the synchronization index $\gamma$ is expected to be closer to 1 than 0. In essence, $\gamma$ measures phase synchronization (or phase locking) between two signals. Finite data length may lead to non-zero values of $\gamma$ even for uncoupled oscillators. But most of our results have synchronization index between 0.1 and 0.6 ($0.1 < \gamma < 0.6$). Within this range, there is a preferred phase locking angle. Thus, though it is not a very rigorous definition, from a practical perspective, the dynamics within this range of $\gamma$ can be practically considered as partially, moderately, and imperfectly synchronized.

We apply the method used in earlier studies of temporal patterning of synchrony in electrophysiological data (Park et al., 2010; Ahn et al., 2011, 2014; Ahn and Rubchinsky, 2013) to characterize the temporal patterning of synchronization. The detailed description of this time-series analysis has been published earlier (Park et al., 2010; Ahn et al., 2011; Rubchinsky et al., 2012); we will briefly summarize the approach below. When $\varphi_1(t_j)$ crosses zero from negative to positive values, we record the value of $\varphi_2(t_j)$ to a set of phase difference $\{\phi_i = \varphi_2(t_j) | \varphi_1(t_j) > 0$ and $\varphi_1(t_{j-1}) < 0 \}$. The values $\phi_i$ cluster around some mean, called preferred phase difference, if the signals are partially synchronized. At a cycle $i$, if $\phi_i$ is more than $\pi/2$ away from the preferred phase difference, the signals are classified as being desynchronized; otherwise, they are synchronized. The duration of desynchronization is computed as the number of consecutive desynchronized cycles.

The modeled network is moderately synchronized; it comes in and out of synchronous state. To characterize this temporal patterning of synchronization, we use the distribution of desynchronization duration. We consider the mode of this distribution (i.e. the most common desynchronization duration), the probability (the relative frequency) of the modal desynchronization duration, the average desynchronization duration, and the so-called desynchronization ratio. A desynchronization ratio is the ratio of the frequency of desynchronization episodes lasting for 1 cycle to the frequency of desynchronization episodes lasting for 5 cycles or more. This characteristic was used in previous experimental studies (Ahn et al., 2014, 2018; Malaia et al., 2020) and the computation study (Nguyen and Rubchinsky, 2021). This quantity informs if the system is biased toward short desynchronization (large ratio) or long desynchronization (small ratio).

Figure 1DE shows an example of dynamics with short desynchronizations. Panel D illustrates a short segment of the time series (sine of phase $\varphi_*(t)$) for each circuit; the demonstration indicates that the time series from two circuits come in and out of synchronization over time. The distribution of desynchronization durations is shown in panel E. It has mode 1 (the most common desynchronization duration lasts 1 cycle of oscillation). The imperfect synchronization dynamics is due to the effects of moderate coupling and heterogeneity (there is no noise considered in this study).

## 3. RESULTS

Here we present the results of our study of the effect of synaptic excitation (section 3.1), synaptic inhibition (sections 3.2), and kinetics of delayed-rectifier potassium channel (section 3.3) on the temporal profile of synchronization in a pyramidal interneuron gamma rhythm network. These parameters are varied independently of each other. The major objective of the study is to show how biologically reasonable network changes may lead to the changes in the temporal patterning of synchronous dynamics. This approach to parameter change allows for addressing this issue without the need to analyze the system in a high-dimensional parameter space (even though the latter would likely reveal more dynamically rich phenomena). Since the connections and the inputs are random, we ran 50 simulations for each combination of parameters. The results are presented as mean values and their standard errors. The ranges of parameters, in each case, are chosen so that the mean firing rate of the networks are around the high 20's Hz or low 30's Hz to the high 30's Hz, so that the ranges are close to the low gamma band.

We also consider the response of networks with different temporal patterning of neural synchrony to periodic input signals (section 3.4).

**3.1 Effect of synaptic excitation on the temporal patterning of intermittent synchronization**

In this section, we describe how the temporal patterns of synchronization are affected by changes in the strength of excitatory synapses. We start with varying the local excitatory connection strength $g_{EI}$ from 0.009 to 0.02. The changes of dynamics properties are provided in the Fig. 2, left column. The mean value of frequency of activity decreases from 34 Hz to 28 Hz. The mean value of synchronization index decreases from 0.25 to 0.11 (Fig. 2A). The distribution of desynchronization durations has a mode of 1 cycle of oscillations in almost all simulations of all cases. The probability of desynchronization duration lasting 1 cycle is shown in Fig. 2B, and it decreases from 0.49 to 0.36, indicating that desynchronizations are becoming longer. Fig. 2B also demonstrates the probability of desynchronization duration lasting 5 cycles or more; this quantity increases from 0.10 to 0.22. The inset shows the probability of desynchronization duration lasting 2 cycles (solid), 3 cycles (dashed), and 4 cycles (dotted); these values stay relative constant. This is illustrated with two histograms, B1 ($g_{EI}$ is small) and B2 ($g_{EI}$ is big). In these two histograms, the probability of desynchronization duration lasting 1 cycle and that of 5 cycles or more change substantially while those of 2 cycles, 3 cycles, and 4 cycles remain relatively similar.

The mean value of average desynchronization duration shows an increasing trend (Fig. 2C), from 2.20 to 3.15 The mean value of desynchronization ratio, in general, has a decreasing trend (Fig. 2D), from 6.53 to 2.29. All these suggest the intermittent synchrony shows longer desynchronizations with the increase of $g_{EI}$.

Variation of inter-network excitatory synapses strength $c_{EI}$ shows the opposite trend, Fig. 2, right column. The mean value of frequency of activity decreases from 34 Hz to 30 Hz when $c_{EI}$ goes from 0 to 0.02. The mean value of synchronization strength (in contrast with $g_{EI}$ variation) substantially increases from 0.04 to 0.24 (Fig. 2A). The distribution of desynchronization durations has a mode of 1 cycle of oscillations for most of the considered values of $c_{EI}$. The mean values of probability of desynchronization duration lasting 1 cycle increases from 0.29 to 0.53 (Fig. 2B) while the mean values of probability of desynchronization duration lasting 5 cycles or more decreases from 0.26 to 0.11 (Fig. 2B). Similar to the case of $g_{EI}$ variation, the mean values of probability of desynchronization duration lasting 2 to 4 cycles do not vary significantly. Histograms B3 and B4 show examples of the distribution of desynchronization duration for small and large values of $c_{EI}$, respectively. The average desynchronization duration shows a decreasing trend from 3.61 to 2.30 (Fig. 2C); the desynchronization ratio generally has an increasing trend from 1.26 to 6.30 (Fig. 2D). Thus, the intermittent synchrony shows shorter desynchronizations with the increase of $c_{EI}$. Overall, the effects of $g_{EI}$ and $c_{EI}$ are opposite: stronger local synapses longer desynchronization durations while stronger distant synapses promote shorter desynchronization durations.

### 3.2 Effect of synaptic inhibition synapse on the temporal patterning intermittent synchronization

#### 3.2.1 Inhibitory to excitatory connections

We vary local connection $g_{IE}$ from 0.3 to 0.4 to examine its role on the temporal dynamics of synchronization, see Fig. 3 left column. In this range, the firing frequency decreases from 34 Hz to 29 Hz. The mean value of synchronization strength decreases mildly from around 0.22 to around 0.17 (Fig. 3A). The relative frequency of desynchronization duration lasting 1 cycle is shown in Fig. 3B; the distribution of desynchronization duration has mode 1 in all cases. The relative frequency of 1 cycle does not show any particular trend, staying around 0.44 to 0.47. Similar, the probability of desynchronization duration lasting 2 cycles, 3 cycles, 4, cycles, and 5 cycles or more stay relatively constant, hovering around 0.10 and 0.15. Histograms B1 and B2 show examples of the distribution of desynchronization duration. Both histograms are relatively similar.

Similarly, the average desynchronization duration (Fig. 3C) does not show a clear trend of either increasing or decreasing, its values vary between 2.42 and 2.75. The desynchronization ratio (Fig. 3D) varies more noticeably between 3.47 and 6.19. In conclusion, there is no clear trend for the temporal dynamics of synchronization observed in the case of $g_{IE}$; relative frequency of cycle 1 and other cycles, average values, and the desynchronization ratio fluctuate in the domain of $g_{IE}$.

The effect of inter-network connection $c_{IE}$ on intermittent synchronization is illustrated in Fig. 3, right column. The frequency decreases from 32 Hz to 30 Hz with $c_{IE}$ for $c_{IE}$ in between 0 to 0.2 The synchronization index increases from 0.14 to 0.39 (Fig. 3A). The mode of the desynchronization distribution is mostly 1. The mean value of the probability of desynchronization duration lasting 1 cycle increases from 0.4 to 0.62, and the mean value of the probability of desynchronization duration lasting 5 cycles or more decreases from 0.18 to 0.04 (Fig. 3B). The mean values of the probability of desynchronization duration lasting 2 cycles and 3 cycles stay relatively constant while that of the 4 cycles shows a mild decreasing trend. Examples of the distribution of desynchronization durations for small $c_{IE}$ and large $c_{IE}$ are shown in histograms B3 and B4, respectively.

Average desynchronization duration decreases, with the mean values going from 2.89 to 1.71 (Fig. 3C). Desynchronization ratio increases, with the mean values going from 2.78 to 11.25 (Fig. 3D). In essence, while change of $g_{IE}$ does not elicit substantial change in the dynamics of desynchronizations, increasing distant connection $c_{IE}$ promotes shorter desynchronization.

#### 3.2.2 Inhibitory to inhibitory connections

Next, we examine the effect of local coupling among inhibitory neurons on pattern of synchronization by varying $g_{II}$ from 0.03 to 0.25 (Fig. 4, left column). The mean firing frequency increases from 26 Hz to 40 Hz. The synchronization index drops drastically from 0.36 to 0.05 for values of $g_{II}$ between 0.03 and 0.14; it then remains small for $g_{II}$ values greater than 0.16 (Fig. 4A). The probability of desynchronization duration lasting 1 cycle depicts a prominent decreasing trend; the mean value goes from 0.50 to 0.32 (Fig. 4B). The probability of desynchronization duration lasting 5 cycles or more increases drastically from 0.07 to 0.32. While the probability of desynchronization duration lasting 3 or 4 cycles stays relatively constant, that of 2 cycles shows a clear decreasing trend from 0.23 to 0.14. Histograms B1 and B2 illustrate desynchronization durations for the cases of small $g_{II}$ and large $g_{II}$.

Figure 4C demonstrates the average desynchronization duration, mean value increasing from 2.14 to 4.08. Figure 4D depicts the change of the desynchronization ratio; this quantity shows an obvious decreasing trend, mean value going from 8.08 to 1.21. In brief, the duration of desynchronizations increases for larger values of $g_{II}$.

While a comprehensive study of the size effect is beyond the scope of our paper, we performed a limited simulation varying $g_{II}$ using 100 neurons and 250 neurons in each circuit. The result presented here is consistent for larger models; we observed that desynch duration increases as $g_{II}$ increases.

The effect of inter-network connections between inhibitory neurons is shown in Fig. 4, right column. The firing rate increases slightly from 30 to 33 Hz when $c_{II}$ is varied from 0 to 0.2. The synchronization index remains relatively constant, mean value fluctuating around 0.2, see Fig. 4A. Similarly to all the cases considered above, the distribution of synchronization has mostly mode 1.

The probability of desynchronization duration lasting 1 cycle depicts a mild increasing trend, mean value going from 0.4 to 0.51, see Fig. 4B. The probability of desynchronization duration lasting 3 or 4 cycles stay relatively constant. On the other hand, the probability of desynchronization duration lasting 2 cycles and that of 5 cycles or more shows a mild decreasing trend, decreasing from 0.22 to 0.18 and from 0.16 to 0.12, respectively. Examples of the distribution of desynchronization distributions for small $c_{II}$ and large $c_{II}$ are shown in histograms B3 and B4. The mean value of average desynchronization duration decreases from 2.72 to 2.33 (Fig. 4C). The mean value of desynchronization ratio increases from 3.82 to 6.31 (Fig. 4D). In conclusion, larger values of $c_{II}$ tend to produce shorter desynchronization dynamics.

**3.3 Effect of current kinetics parameters on the temporal patterning of intermittent synchronization**

Earlier modeling studies of very small networks of excitatory coupled Morris Lecar neurons (Ahn and Rubchinsky, 2017) and of pyramidal neurons and interneurons (Nguyen and Rubchinsky, 2021) demonstrated that the timescales of the potassium delayed-rectifier current kinetic can alter the temporal dynamics of synchronization. While it is not clear if this kinetics can be easily altered in an experiment, we aim to see if similar effects exist in a more biologically realistic network investigated in the present study.

For each neuron, the activation time constant of potassium channel is rewritten as:

$$\tau_n(V) = \frac{1}{\varepsilon} \frac{1}{\alpha_n(V) + \beta_n(V)} \tag{14}$$

where $\alpha_n(V)$ and $\beta_n(V)$ are the opening and closing function of potassium channel (see Methods, Equations 8 and 9). Here, parameter $\varepsilon$ changes the peak value of the voltage-dependent activation time constant $\tau_n(V)$. When $\varepsilon$ is small, the activation of the potassium current is slower. The voltage waveform has characteristically spiky shape, and the underlying dynamical system is a relaxation oscillator. When $\varepsilon$ is large, the activation of the potassium current is faster. The waveform becomes more sinusoidal, and the oscillator is more harmonic.

The impact of $\varepsilon$ on the synchronization dynamics is presented in the left column of Fig. 5. When $\varepsilon$ is varied from 0.28 to 2.96, the firing frequency varies from 28 Hz to 38 Hz. Figure 5A (left) shows the effect of $\varepsilon$ on synchronization index, which is decreasing from 0.30 to 0.01. The mode of desynchronization durations distribution is mostly 1. The probability of desynchronization duration lasting 1 cycle decreases from 0.45 to 0.30 while the probability of desynchronization duration lasting 5 cycles or more increases from 0.12 to 0.36. It is noteworthy that the latter probability is bigger than the former probability for values of $\epsilon > 2.6$ (Fig. 5B). Moreover, while probability of desynchronization duration lasting for 3 or 4 cycles remain relatively constant, that of 2 cycles first increases from 0.21 to 0.24 (for $\epsilon$ values between 0.28 and 0.82) and then decreases to 0.12 (when $\epsilon > 0.82$). Histograms of desynchronization durations for some cases of $\epsilon$ are shown in histograms B1 and B2.

Furthermore, the average desynchronization duration increases from 2.46 to 4.39 (Fig. 5C), and the desynchronization ratio decreases from 4.46 to 0.87 (Fig. 5D). In essence, small values of $\varepsilon$ (which produce relaxation oscillations and more spiky waveforms) promote short desynchronizations while large values of $\varepsilon$ promote longer desynchronizations.

Following the aforementioned modeling studies in small networks, we also vary the width of voltage-dependence of the activation time-constant of potassium current. We parametrize $\alpha_n(V)$ and $\beta_n(V)$ as:

$$\alpha_n(V) = \frac{0.032(V+52)}{1-\exp\left(-\frac{1}{\delta}\frac{v+52}{5}\right)} \quad ; \quad \beta_n(V) = 0.5\exp\left(-\frac{1}{\delta}\frac{V+57}{40}\right) \tag{15}$$

for excitatory neurons, and

$$\alpha_n(V) = \frac{0.05(V+34)}{1-\exp\left(-\frac{1}{\delta}\frac{v+34}{10}\right)} \quad ; \quad \beta_n(V) = 0.625\exp\left(-\frac{1}{\delta}\frac{V+44}{80}\right) \tag{16}$$

for inhibitory neurons. Increasing the value of $\delta$ increases the range of voltages where voltage-dependent activation time constant $\tau_n(V)$ is large, thus effectively slowing the current.

The impact of $\delta$ on synchronization dynamics is presented in the right column of Fig. 5. We vary $\delta$ from 0.1 to 1.2, and the frequency varies first goes from 38 Hz to 30 Hz ($0.1 < \delta < 0.6$) and then increases to 34 Hz ($\delta > 0.6$). The synchronization index increases from 0.09 to 0.26 (Fig. 5A). The mode of desynchronization durations distribution is mostly 1. The probability of desynchronization duration lasting 1 cycle slightly increases from 0.40 to 0.47 (Fig. 5B). Meanwhile, the probability of desynchronization duration lasting 5 cycles or more decreases from 0.22 to 0.13, and the probabilities of desynchronization duration lasting 2, 3, and 4 cycles do not vary significantly. Histograms B3 and B4 are examples of desynchronization durations distribution for different cases of $\delta$.

As expected, the average desynchronization duration decreases from 3.17 to 2.40 (Fig. 5C), and the desynchronization duration ratio increases from 2.47 to 5.68 (Fig. 5D). Thus, small values of $\delta$ (and effectively faster current kinetics) promote longer desynchronizations while large values of $\delta$ promote shorter desynchronizations.

While the changes in kinetic properties may not be altered easily in experiments, it is worthy to point out that the changes in the activation time of the potassium channel and the width of voltage-dependence of the activation time constant of potassium current presented here separate the time scales of a typical fast-slow system. In these cases when $\epsilon$ is small (lower value for peak activation time constant) or when $\delta$ is large (larger width of voltage dependence of the activation time constant), there is effectively a delay in the activation of the potassium current. Thus, the network generates more spiky waveforms (i.e. relaxation oscillator). In this relaxation regime, the cellular mechanism promotes synchronization much more rapidly than the cellular mechanism in the sinusoidal waveform regime (probably similarly to Somers and Kopell, 1993). Our simulation shows that the relaxation regime indeed promotes short desynchronizations in this system.

### 3.4 Different temporal patterning of synchronized activity leads to variation in network sensitivity to input signal

Ahn and Rubchinsky (2017) demonstrated that a network with short desynchronization durations may be more sensitive to input as compared to the network with longer desynchronization durations (even if the average synchronization strength and frequency are similar). This may be considered as a potential functional advantage of short desynchronizations dynamics observed in experiments (see references in Introduction). In this section, we consider examples in which different temporal patterning of synchronized dynamics in our PING network may alter the response of the network to periodic input signals.

We impose a common synaptic input into each network of our model; this input is supplied to each excitatory neuron in both circuits. This synaptic input is coming from an excitatory neuron that has the same form as the excitatory neurons in the networks, see Methods. The connection strength between the external neuron and each excitatory neuron from the two circuits is $G$. The diagram of this network arrangement is presented in the Fig. 6A.

We simulate the above circuit (Fig. 6A) in two different scenarios. In the first scenario, the synaptic weights of stimulated network are selected in such a way that the distribution of desynchronization durations of the original model (no common synaptic input, $G = 0$) has mode 1. We call it "mode 1 system" or short desynchronization system, see Fig. 6B (top). In the second scenario, synaptic weights of the stimulated network are selected in such a way that the distribution of desynchronization durations of the original model (no common synaptic input, $G = 0$) has mode 2. We call it "mode 2 system" or long desynchronization system, see Fig. 6B (bottom). "Mode 1" system has $c_{IE} = 0.1$, and "mode 2 system" has $c_{IE} = 0.18$. The other parameters are the same between scenarios. While these two systems exhibit different temporal patterns of desynchronization, they have similar synchronization index (0.3 for "mode 1 system" and 0.32 for "mode 2 system") and firing frequency (32 Hz for "mode 1 system" and 31 Hz for "mode 2 system"). The common neuronal input is tuned so that its firing frequency is similar to the average firing frequency of the original network (no common synaptic input, $G = 0$). Thus, we essentially compare how networks with different patterning of synchronization (but similar frequency and synchrony strength) respond to input signal.

The input connection strength $G$ is varied from 0 to $0.625 \times 10^{-2}$. As $G$ increases, the two networks become more synchronized, but to a different degree in different scenarios. The synchronization index is denoted as $\gamma_1$ for "mode 1 system" and as $\gamma_2$ for "mode 2 system." Figure 6C draws the normalized difference of synchronization indices,

$$\Delta = \frac{\gamma_1 - \gamma_2}{\gamma_2} \qquad (17)$$

between the two values as a function of input strength. With no input, "mode 2 system" is a bit more synchronized than "mode 1 system". For small nonzero input strength, the normalized difference $\Delta$ is even more negative. This means "mode 2 system" is more synchronized with the same input strength; in other words, "mode 2 system" is more easily synchronizable. For intermediate values of input strength ($0.13 \times 10^{-2} < G < 0.37 \times 10^{-2}$), the normalized difference $\Delta$ is positive. Thus "mode 1 system" is more synchronized than "mode 2 system" in this region even though the input strength to both systems is the same. Finally, the normalized difference $\Delta$ is around 0 when the input strength is large ($G > 0.37 \times 10^{-2}$). In this region, both systems are highly synchronized due to strong input, so that $\gamma_1$ and $\gamma_2$ cannot differ much. In essence, "mode 1 system" and "mode 2 system" vary in the ability to synchronize across different values of input strength.

## 4. DISCUSSION

### 4.1 Synaptic influence on the temporal patterning of synchronization

This study used a medium-sized pyramidal interneuron gamma rhythm (PING) network to investigate the effect of synaptic connection strength on the temporal patterning of partially synchronized gamma oscillations. This partially synchronized regime is (expectedly) observed for intermediate values of synaptic strength. We found that changing synaptic strength does not only change the average synchronization index but also alters the temporal patterning of synchronization (and these two do not necessarily co-vary in the same way). These observations fit with a prior analysis of dynamics in a very small network of excitatory and inhibitory neurons (Nguyen and Rubchinsky, 2021). However, the small number of neurons in that study may raise concerns of biological realism (Borgers et al., 2012). In fact, the previous model used a very small network (2 excitatory and 2 inhibitory neurons in each circuit). Physiologically realistic gamma rhythms may break down if the numbers of cells are too small, and this value depends on the level of heterogeneity, randomness, and the synaptic strength of the network (Borgers et al., 2012). Here we use a medium-sized network and use a biologically plausible ratio of excitatory neurons over inhibitory neurons (4:1 here, and in general tend to be in the 3:1-9:1 range for mammalian cortex).

Moreover, the connection structure used in (Nguyen and Rubchinsky, 2021) is practically one specific connectivity structure, while the results presented here are aggregated across a number of network samples.

The effect of synaptic strength on the temporal patterning of synchronization of neural activity is observed to follow a general trend: stronger local connection ($g_{EI}$ and $g_{II}$) tends to produce longer desynchronization dynamics while stronger distant connection ($c_{EI}$, $c_{IE}$, and $c_{II}$) tends to produce shorter desynchronization dynamics. In short, local synapses (except the case of $g_{IE}$ in which there is no clear observable trend) and distant synapses have the opposite effects on the temporal dynamics of synchronization. Earlier study in a minimal network model (Nguyen and Rubchinsky, 2021) yielded less consistent results and did not exhibit some of these trends. There may be many reasons for this, in particular, smaller network may have somewhat different (and less realistic) mechanisms (combination of heterogeneity, randomness, and synaptic strength) of the gamma rhythmicity. However, the common observation is that the temporal patterning of the synchronized activity can be altered by the variation of synaptic strength (of both local and long-range connections). And with the results of the present analysis of ensemble of medium sized network, it is likely that local synapses and long-range connections in cortical networks may have opposite effect on the temporal dynamics of synchronization regardless of connection structures.

**4.2 Prevalence of short desynchronizations**

While the temporal patterns of synchronized dynamics show a lot of variability depending on the parameter values, in most of the cases considered, the most prevalent desynchronization duration is one cycle of oscillations. The number of desynchronized intervals and the relative frequency of short and long desynchronizations may vary substantially, and the average synchronization strength may vary too. Nevertheless, the mode of the distribution of desynchronization durations is almost always 1; the probability of desynchronization duration lasting 1 cycle is consistently larger than that of higher cycles. Averaging across simulations, the probability of desynchronization duration of shorter cycles is always bigger than that of higher cycles (prob. Cycle 1 > prob. Cycle 2 > prob. Cycle 3 > prob. Cycle 4 > prob. Cycle 5 or more), see panel B of Figs. 2-5 (even though this is not necessarily true for some arbitrary partially synchronized systems, Ahn et al., 2011, Ahn and Rubchinsky, 2017). It is noteworthy to point out that as network alters its temporal dynamics from short desynchronization to long desynchronization or vice versa, the changes in the relative frequency of desynchronization of cycle 1 and that of cycle 5 or more are generally more noticeable than the changes in other cycles (cycles 2-4), see panel B of Figs. 2-5.

This prevalence of short desynchronizations fits well with experimental studies of temporal synchrony patterns in the brain, which seem to indicate that prevalent short desynchronizations may be very common (if not universal) for neural synchronization in the brain. Short desynchronizations were observed in alpha and beta band at rest and during motor task (Ahn and Rubchinsky, 2013), in theta oscillations in drug addicted and in control state (Ahn et al., 2014), in beta oscillations in Parkinson's disease (Park et al., 2010; Ratnadurai-Giridharan et al., 2016; Ahn et al., 2018; Dos Santos Lima et al., 2020), and in theta, beta, and low frequency gamma in autism spectrum disorder (Malaia et al., 2020). It was found in different brain areas (different parts of the cortex and basal ganglia), in different health status, in different types of recorded electrophysiological signals (EEG, LFP, spiking units). Generic synchronized oscillators can easily exhibit longer desynchronizations (Ahn et al., 2011; Ahn and Rubchinsky, 2017), but PING networks considered here show predominantly short desynchronizations without any special tuning.

Numerical studies in a minimal network of two neurons suggested that slow kinetics of delayed-rectifier potassium current (which creates relaxation oscillator and characteristically sharp waveform of spike) promotes short desynchronizations dynamics (Ahn and Rubchinsky, 2017). Variation of parameters defining this current kinetics in the model considered here has similar effect (see Section 3.3); however, short desynchronizations dynamics in the present model is much more robust than in (Ahn and Rubchinsky, 2017). While other mechanisms of short desynchronizations, such as synaptic plasticity (Zirkle and Rubchinsky, 2020) and channel noise (Zirkle and Rubchinsky, 2021), have been suggested, the network considered here exhibits short desynchronization dynamics without any plasticity or noise, signifying that those mechanisms are likely not a necessary condition for short desynchronization.

Short desynchronization dynamics was conjectured in (Anh and Rubchinsky, 2013) to be functionally beneficial for neural circuits because it may facilitate quick formation and dissipation of transient neuronal assemblies (as the transiently synchronized states are already created and dissolved in a quick and dynamic way in the case of short desynchronizations). It may be interesting to see if the modeling framework employed here will be possible to use for exploring this hypothesis.

### 4.3 Network synchronizability for different temporal patterning of synchronization

An earlier study suggested that different temporal patterns of synchronization may have an influence on how neural circuits synchronize with input (Ahn and Rubchinsky, 2017). To show that the temporal patterning of synchronization may affect potential function of PING network, we considered how the system used here can respond to external synchronizing input signal (see Section 3.4). We showed that model circuits with the same base level of synchrony will respond to synaptic input in different ways, depending on the temporal pattern of desynchronizations in the circuits without input. The same strength of synaptic input in the gamma frequency range may elevate synchrony in a circuit with shorter desynchronizations by a larger amount than in a circuit with longer desynchronizations (even though both circuits have the same synchrony strength before the input is introduced). Therefore, the way how synchronized and desynchronized intervals are distributed in time may make a difference for how a circuit responds to incoming signals. This suggests that the issue of how temporal patterning of synchrony may affect synchronizability is worthy of detailed investigation. The results presented here can lay a foundation for further computational as well as experimental studies on the relationship between the temporal synchronization dynamics and neuronal function.

### 4.4 Potential relationship with experimental observations

The impact of synaptic strength on the temporal patterning of gamma band synchronization is interesting to consider in the context of the synaptic mechanisms of gamma rhythm and synaptic effects of gamma in health and disease. Synaptic connections between inhibitory neurons and excitatory neurons are known to be important in mediating spike synchronization in gamma band (e.g., Buzsaki 2012; Borger 2017; Salkoff et. al., 2015). Gamma band activity is associated with many brain functions (Fell and Axmacher, 2011; Colgin, 2011; Buzsaki and Schomburg, 2015; Fries, 2015). So, it is natural to see that brain disorders associated with pathologies of gamma rhythm may be driven by abnormalities of synaptic strength. Thus, schizophrenia has been associated with abnormalities in synchronization of gamma oscillations (Uhlhaas and Singer, 2010, 2015; Spellman and Gordon, 2015; Pittman-Polletta et al., 2015), which can result from abnormalities of synaptic strength, including changes in synaptic inhibition and excitatory-inhibitory balance (Lewis et al., 2005; Vierling-Claassen, et al., 2008; Lisman, 2012; Murray et al., 2014; Grent-'t-Jong et al., 2018).

Our results show that synaptic changes may affect not only average synchrony strength, but also the temporal patterning of synchronization. It was suggested (Ahn and Rubchinsky 2013, 2017, see discussion above) that even if the synchrony strength is not changed, the fine temporal structure of how neural system gets in and out of synchronized state may have implications for information processing. The present study's observation of networks with different temporal dynamics of synchronization having different sensitivity to common synaptic input provides one potential mechanism for it. Thus, the synaptic changes, which affect gamma oscillations and result in different properties of information processing in the brain (and its abnormalities in several neurological and neuropsychiatric disorders) may mediate physiological properties of neural circuits not only via change in the average synchrony level, but also via change of how synchrony is patterned in time over very short time scales.

**Funding**

The study was supported by NSF DMS 1813819.

**References**

Ahn, S., and Rubchinsky, L. L. (2013). Short desynchronization episodes prevail in synchronous dynamics of human brain rhythms. *Chaos* 23: 013138.

Ahn, S., and Rubchinsky, L. L. (2017). Potential mechanisms and functions of intermittent neural synchronization. *Front. Comput. Neurosci*. 11: 44.

Ahn, S., Park, C., and Rubchinsky, L. L. (2011). Detecting the temporal structure of intermittent phase locking. Phys Rev E Stat Nonlin Soft Matter Phys 84: 016201.

Ahn, S., Rubchinsky, L. L., and Lapish, C. C. (2014). Dynamical reorganization of synchronous activity patterns in prefrontal cortex - hippocampus networks during behavioral sensitization. *Cereb. Cortex* 24: 2553-2561.

Ahn, S., Zauber, S. E., Witt, T., Worth, R. M., and Rubchinsky, L. L. (2018). Neural synchronization: Average strength vs. temporal patterning. *Clin. Neurophysiol*. 129: 842-844.

Bansal, K., Garcia, J. O., Tompson, S. H., Verstynen, T. Vettel, J. M., and Muldoon, S. F. (2019). Cognitive chimera states in human brain networks. Sci. Adv. 5:eaau8535.

Borgers, C. (2017). *An Introduction to Modeling Neuronal Dynamics*. Springer International Publishing.

Borgers, C., Franzesi, G.T., Lebeau, F. E. N., Boyden, E., and Kopell, N. (2012). Minimal size of cell assemblies coordinated by gamma oscillations. *PLoS Comp. Biol*. 8: e1002362.

Borgers C, Li J, Kopell N. (2014) Approximate, not perfect synchrony maximizes the downstream effectiveness of excitatory neuronal ensembles. J Math Neurosci. 4:10.

Buzsáki, G. (2006) *Rhythms of the Brain*. Oxford University Press.

Buzsáki, G., and Schomburg, E. W. (2015). What does gamma coherence tell us about inter-regional neural communication? Nat. Neurosci. 18, 484–489.

Buzsáki G, Anastassiou CA, Koch C. (2012) The origin of extracellular fields and currents--EEG, ECoG, LFP and spikes. Nat Rev Neurosci. 13:407-420.

Buzsáki, G., and Wang, X. J. (2012). Mechanisms of gamma oscillations. *Annu Rev neurosci* 35: 203–225.

Chouzouris, T., Omelchenko, I., Zakharova, A., Hlinka, J., Jiruska, P., and Scholl, E. (2018). Chimera states in brain networks: empirical neural vs. modular fractal connectivity. Chaos 28:045112.

Colgin, L. L. (2011). Oscillations and hippocampal-prefrontal synchrony. Curr. Opin. Neurobiol. 21, 467–474.

Dos Santos Lima, G. Z., Targa, A. D. S., de Freitas Cavalcante, S., Rodrigues, L. S., Fontenele-Araujo, J., Torterolo, P., Andersen, M. L., and Lima, M. M. S. (2020) Disruption of neocortical synchronisation during slow-wave sleep in the rotenone model of Parkinson's disease. *J Sleep Res*, Aug 31: e13170. doi: 10.1111/jsr.13170. Online ahead of print.

Dumont, G. and Gutkin, B. (2019) Macroscopic phase resetting-curves determine oscillatory coherence and signal transfer in inter-coupled neural circuits. *PLoS Comp. Biol*. 15(5): e07019.

Ementrout, G. B., and Kopell, N. (1998). Fine structure of neural spiking and synchronization in the presence of conduction delays. *Proc. Nat. Acad. Sci. USA* 95: 1259–1264.

Fell, J., and Axmacher, N. (2011). The role of phase synchronization in memory processes. *Nat. Rev. Neurosci*. 12, 105–118.

Fries, P. (2015). Rhythms for cognition: communication through coherence. *Neuron* 88, 220–235

Hurtado, J. M., Rubchinsky, L. L., and Sigvardt, K. A. (2004). Statistical method for detection of phase-locking episodes in neural oscillations. *J Neurophysiol* 91: 1883-98.

Kanagaraj, S., Moroz, I., Durairaj, P. et al. Imperfect chimera and synchronization in a hybrid adaptive conductance based exponential integrate and fire neuron model. Cogn Neurodyn 18, 473–484 (2024).

Lewis, D. A., Hashimoto, T., and Volk, D. W. (2005). Cortical inhibitory neurons and schizophrenia. *Nat Rev Neurosci,* 6: 312-324.

Lisman, J. (2012). Excitation, inhibition, local oscillations, or large-scale loops: what causes the symptoms of schizophrenia? *Curr Opin Neurobiol*. 22: 537-544.

Malaia, E., Ahn, S., and Rubchinsky, L. L. (2020). Dysregulation of temporal dynamics of synchronous neural activity in adolescents on autism spectrum. *Autism Res* 13: 24-31.

Nguyen, Q.-A. and Rubchinsky, L.L. (2021) Temporal patterns of synchrony in a pyramidal-interneuron gamma (PING) network. *Chaos* 31: 043134.

Palmigiano, A., Geisel, T., Wolf, F., Battaglia, D. (2017) Flexible information routing by transient synchrony. Nat Neurosci. 20:1014-1022.

Park, C., Worth, R. M., and Rubchinsky, L. L. (2010). Fine temporal structure of beta oscillations synchronization in subthalamic nucleus in Parkinson's disease, *J Neurophysiol.* 103:2707-2716.

Pikovsky, A., Rosenblum, M., and Kurths, J. (2001). *Synchronization: A Universal Concept in Nonlinear Sciences.* Cambridge: Cambridge University Press.

Quax, S., Jensen, O., and Tiesinga, P. (2017) Top-down control of cortical gamma-band communication via pulvinar induced phase shifts in the alpha rhythm. *PLoS Comp. Biol.* 13(5): e1005519

Ratnadurai-Giridharan, S., Zauber, S. E., Worth, R. M., Witt, T., Ahn, S., and Rubchinsky, L. L. (2016). Temporal patterning of neural synchrony in the basal ganglia in Parkinson's disease. *Clin Neurophysiol* 127: 1743–1745.

Rubchinsky, L. L., Park, C., and Worth, R. M. (2012). Intermittent neural synchronization in Parkinson's disease. *Nonlinear Dyn* 68: 329–346.

Sahara, S., Yanagawa, Y., O'Leary, D., and Stevens, C. (2012). The fraction of cortical GABAergic neurons is constant from near the start of cortical neurogenesis to adulthood. *J Neurosci* 32(14): 4755-4761.

Salkoff, D., Zagha, E., Yüzgeç, Ö., and McCormick, D. (2015). Synaptic Mechanisms of Tight Spike Synchrony at Gamma Frequency in Cerebral Cortex. *J Neurosci* 35: 10236-10251.

Somers, D. and Kopell, N. (1993). Rapid Synchronization through Fast Threshold Modulation. Biological Cybernetics 68, 393–407. doi: 10.1007/BF00198772

Stimberg M, Brette R, Goodman DFM (2019). Brian 2, an intuitive and efficient neural simulator. *eLife* 8:e47314. doi: 10.7554/eLife.47314

Sun L, Grützner C, Bölte S, Wibral M, Tozman T, Schlitt S, Poustka F, Singer W, Freitag CM, Uhlhaas PJ. (2012) Impaired Gamma-Band Activity during Perceptual Organization in Adults with Autism Spectrum Disorders: Evidence for Dysfunctional Network Activity in Frontal-Posterior Cortices. *J Neurosci*. 32: 9563-9573.

Traub, R. D., and Miles, R. (1991). *Neuronal Networks of the Hippocampus*, Cambridge, University Press, Cambridge, UK.

Uhlhaas, P. J., and Singer, W. (2010). Abnormal neural oscillations and synchrony in schizophrenia. *Nat. Rev. Neurosci.* 11: 100–113.

Uhlhaas P. J., and Singer, W. (2015) Oscillations and neuronal dynamics in schizophrenia: the search for basic symptoms and translational opportunities. Biol Psychiatry 77:1001-1009.

Vierling-Claassen, D., Siekmeier, P., Stufflebeam, S., and Kopell, N. (2008) Modeling GABA alterations in schizophrenia: a link between impaired inhibition and altered gamma and beta range auditory entrainment. *J Neurophysiol*, 99: 2656-2671.

Wang, X. J., and Buzsáki, G. (1996). Gamma oscillation by synaptic inhibition in a hippocampal interneuronal network model. *J. Neurosci*. 16: 6402–6413.

Zirkle, J, and Rubchinsky, L. L. (2020) Spike-timing dependent plasticity effect on the temporal patterning of neural synchronization. *Front Comput Neurosci* 14: 52.

Zirkle, J, and Rubchinsky, L. L. (2021) Noise Effect on the Temporal Patterns of Neural Synchrony. *Neural Netw* 141: 30-39.

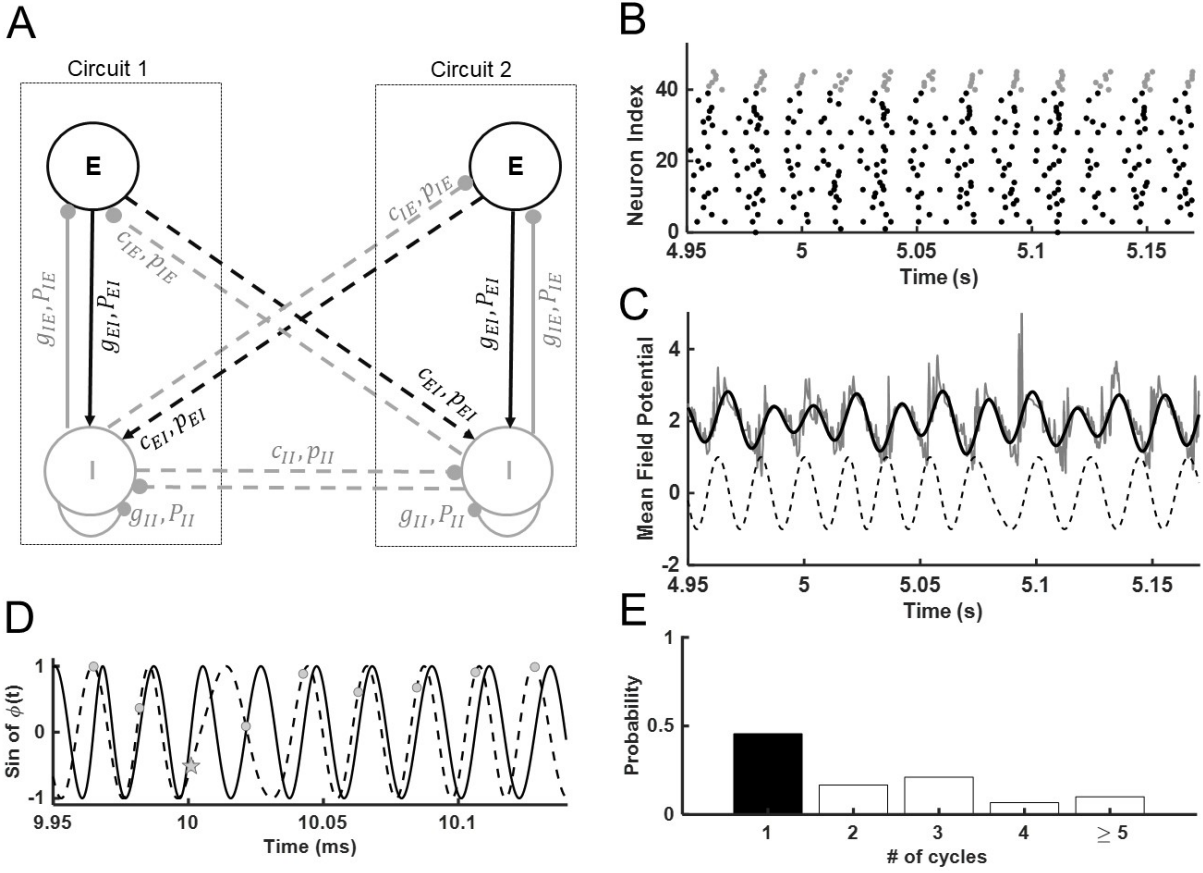

Fig 1: Schematic and the activity of the network. (A) Full model circuitry with 2 individual circuits. Each circuit has an excitatory population E and an inhibitory population I. Excitatory connections ($g_{EI}, g_{EE}, c_{EI}, c_{EE}$) are portrayed with black arrows. Inhibitory connections ($g_{IE}, g_{II}, c_{IE}, c_{II}$) are portrayed with gray arrows. Solid and dotted arrows indicate within-network connection $g_*$ and cross-network connection $c_*$ respectively. The random connectivity matrix has probability $P_*$ for within-network connections and $p_*$ for cross-network connections. (B) Raster plot of a network activity in one network from a particular simulation using the default parameter values. Each dot stands for a spike, black dots are for excitatory neurons, and grey dots are for inhibitory neurons. (C) Local field potential of the network activity depicted in (B) is shown in gray line; the filtered curve of the field potential is shown in black line; the sine of the phase of the field potential is shown in dotted black line. (D) sin of $\phi(t)$, where $\phi(t)$ is the phase of the field potential, for circuit 1 (solid curve) and circuit 2 (dotted curve). Circles and stars indicate the values of the sine of phase of one signal when the phase of the other signal crosses zero from below (see Methods, circles are synchronized states and stars are desynchronized states). (E) The histogram of desynchronization durations of the dynamics of the network in (B); the mode is 1 (black bin).

## A Synchronization Index

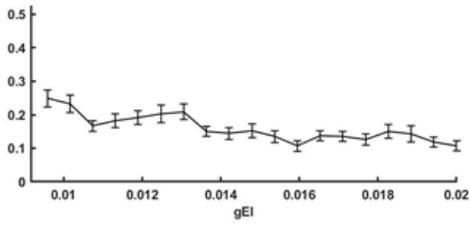
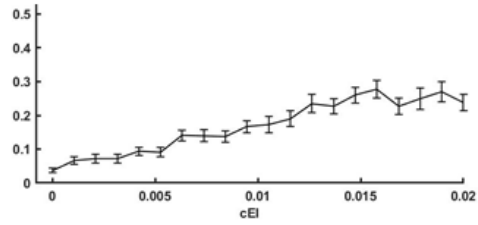

## B Probability of Desynchronization Cycles

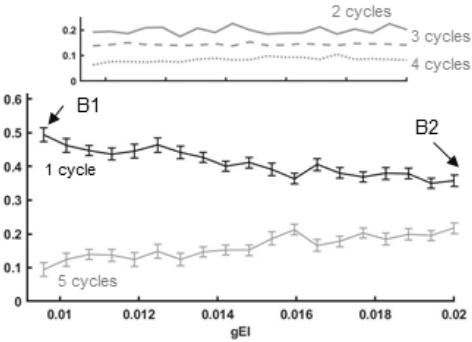
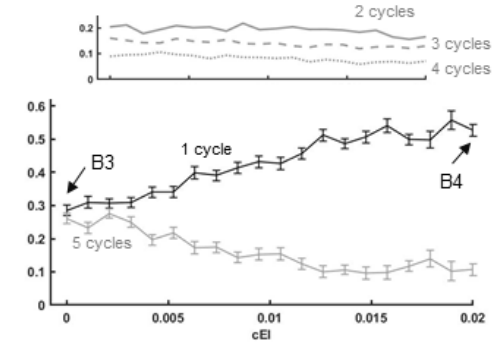

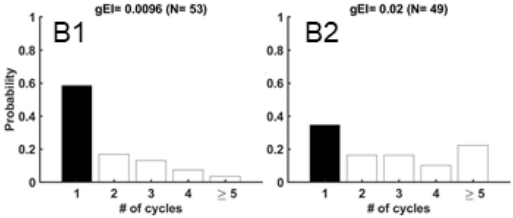
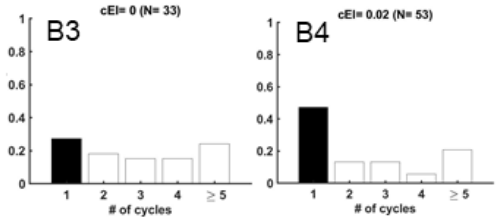

## C Average Desynchronization Duration

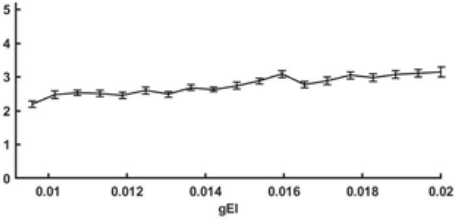
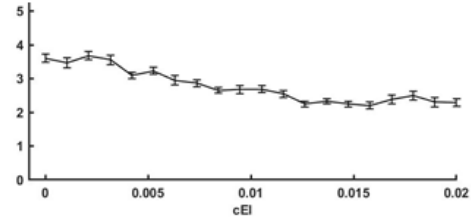

## D Desynchronization Ratio

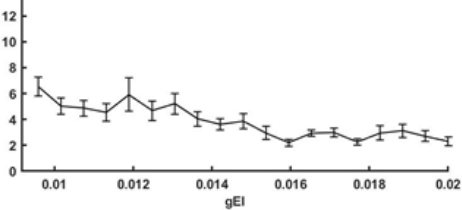
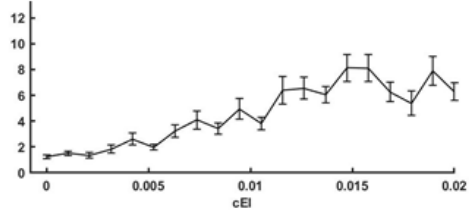

Figure 2: Effects of EI connections on temporal patterning of synchronization. Variation of intra-network excitatory connections $g_{EI}$ and inter-network excitatory connections $c_{EI}$ are presented in the left and right columns, respectively. The results are shown as mean values across 50 simulations. The error bar is the standard error of the mean. (A) Synchronization strength index as a function of $g_{EI}$ and $c_{EI}$. (B) Relative frequency of desynchronization cycles as function of $g_{EI}$ and $c_{EI}$. The histograms (B1, B2, B3, and B4) show examples of desynchronization distributions for the cases with longer and shorter desynchronizations. The black bin is the modal bin. The arrows indicate the parameter values of the examples. (C) Average desynchronization duration as a function of $g_{EI}$ and $c_{EI}$. (D) Desynchronization ratio as a function of $g_{EI}$ and $c_{EI}$.

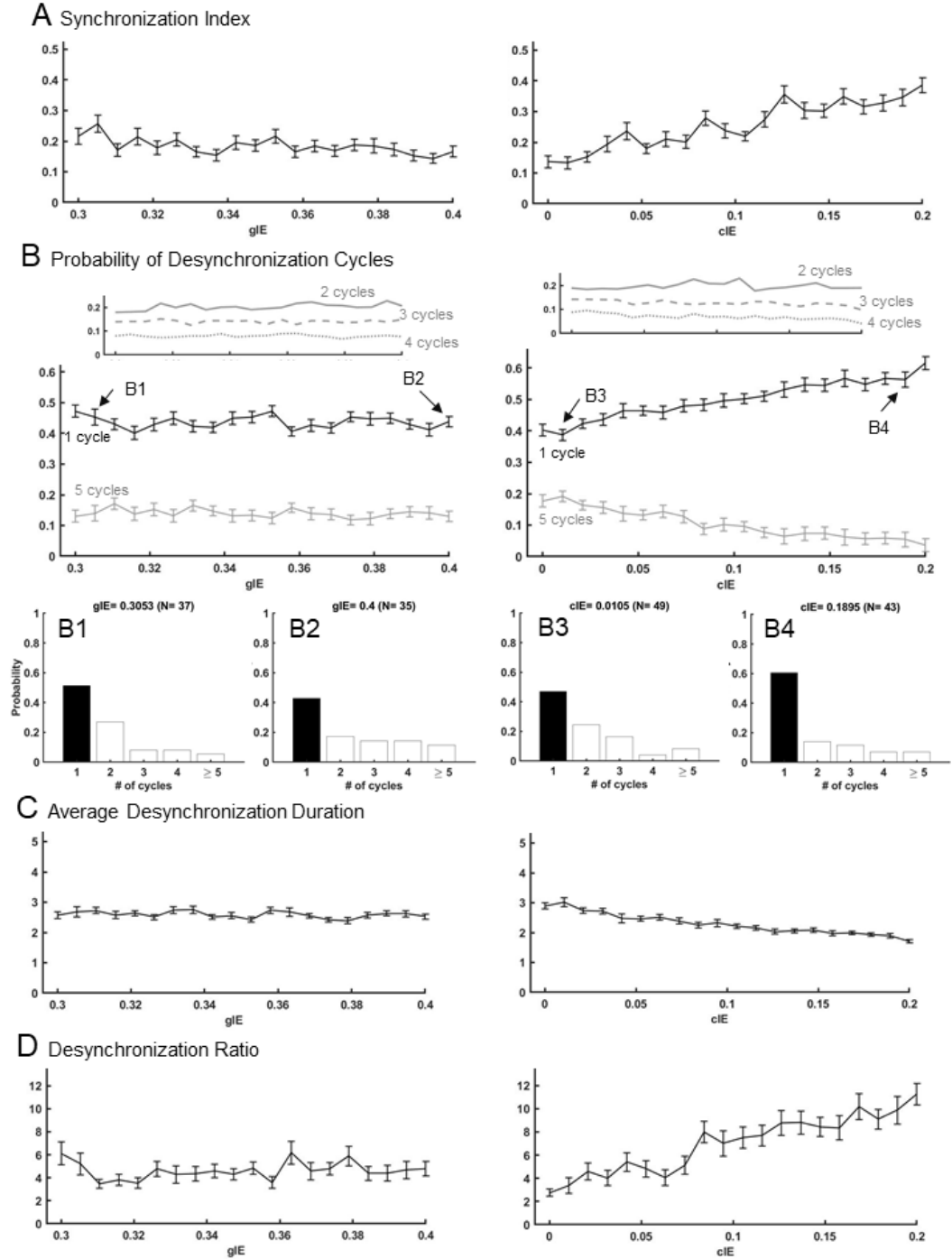

Figure 3: Effects of IE connections on temporal patterning of synchronization. Variation of intra-network inhibitory-to-excitatory connections $g_{IE}$ and inter-network inhibitory-to-excitatory connections $c_{IE}$ are presented in the left and right columns, respectively. The results are shown as mean values across 50 simulations. The error bar is the standard error of the mean. (A) Synchronization strength index as a function of $g_{IE}$ and $c_{IE}$. (B) Relative frequency of desynchronization cycles as function of $g_{IE}$ and $c_{IE}$.. The histograms (B1, B2, B3, and B4) show examples of desynchronization distributions for the cases with longer and shorter desynchronizations. The black bin is the modal bin. The arrows indicate the parameter values of the examples. (C) Average desynchronization duration as a function of $g_{IE}$ and $c_{IE}$. (D) Desynchronization ratio as a function of $g_{IE}$ and $c_{IE}$.

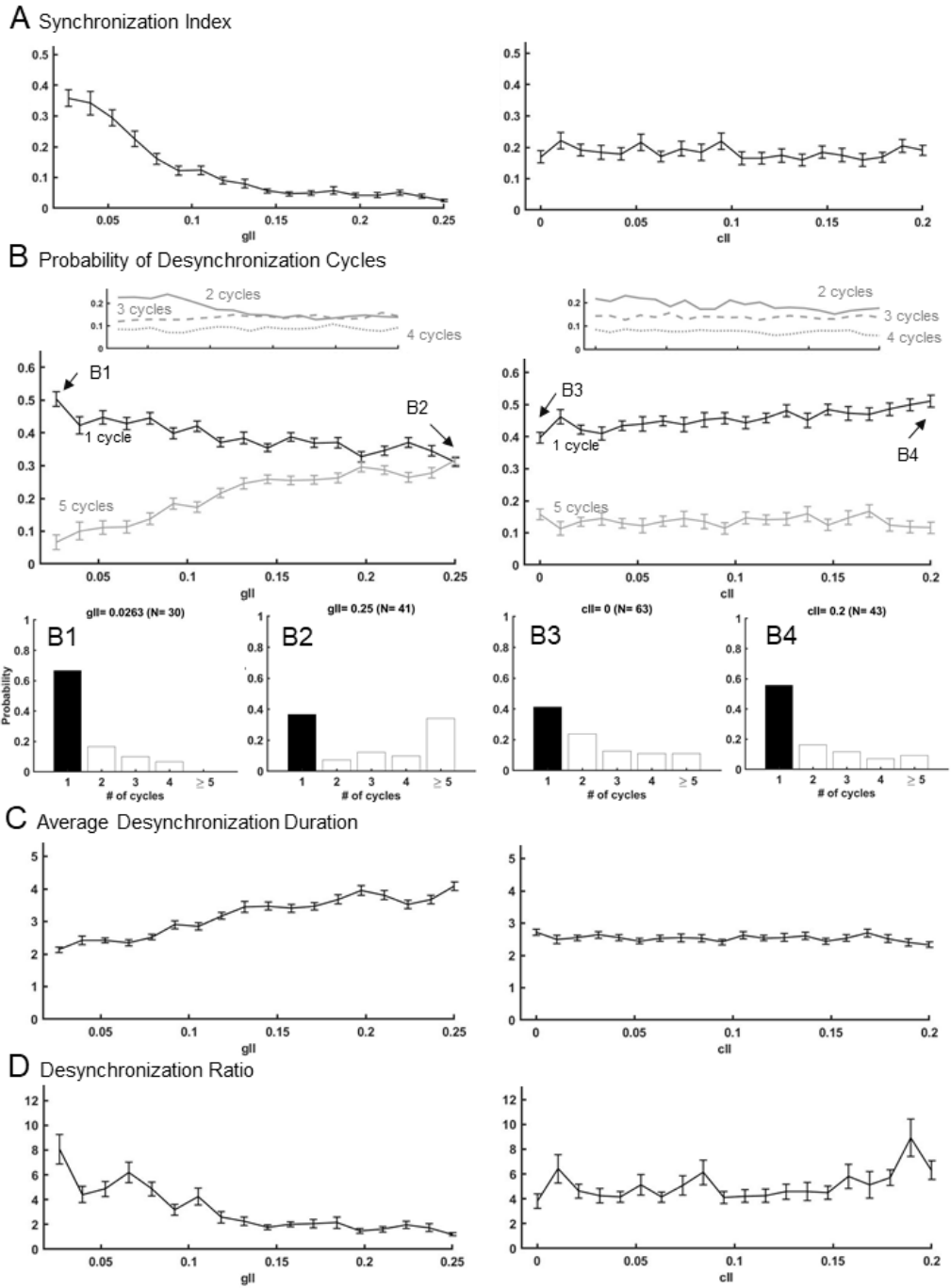

Figure 4: Effects of II connections on temporal patterning of synchronization. Variation of intra-network inhibitory-to-inhibitory connections $g_{II}$ and inter-network inhibitory-to-inhibitory connections $c_{II}$ are presented in the left and right columns, respectively. The results are shown as mean values across 50 simulations. The error bar is the standard error of the mean. (A) Synchronization strength index as a function of $g_{II}$ and $c_{II}$. (B) Relative frequency of desynchronization cycles as function of $g_{II}$ and $c_{II}$. The histograms (B1, B2, B3, and B4) show examples of desynchronization distributions for the cases with longer and shorter desynchronizations. The black bin is the modal bin. The arrows indicate the parameter values of the examples. (C) Average desynchronization duration as a function of $g_{II}$ and $c_{II}$. (D) Desynchronization ratio as a function of $g_{II}$ and $c_{II}$.

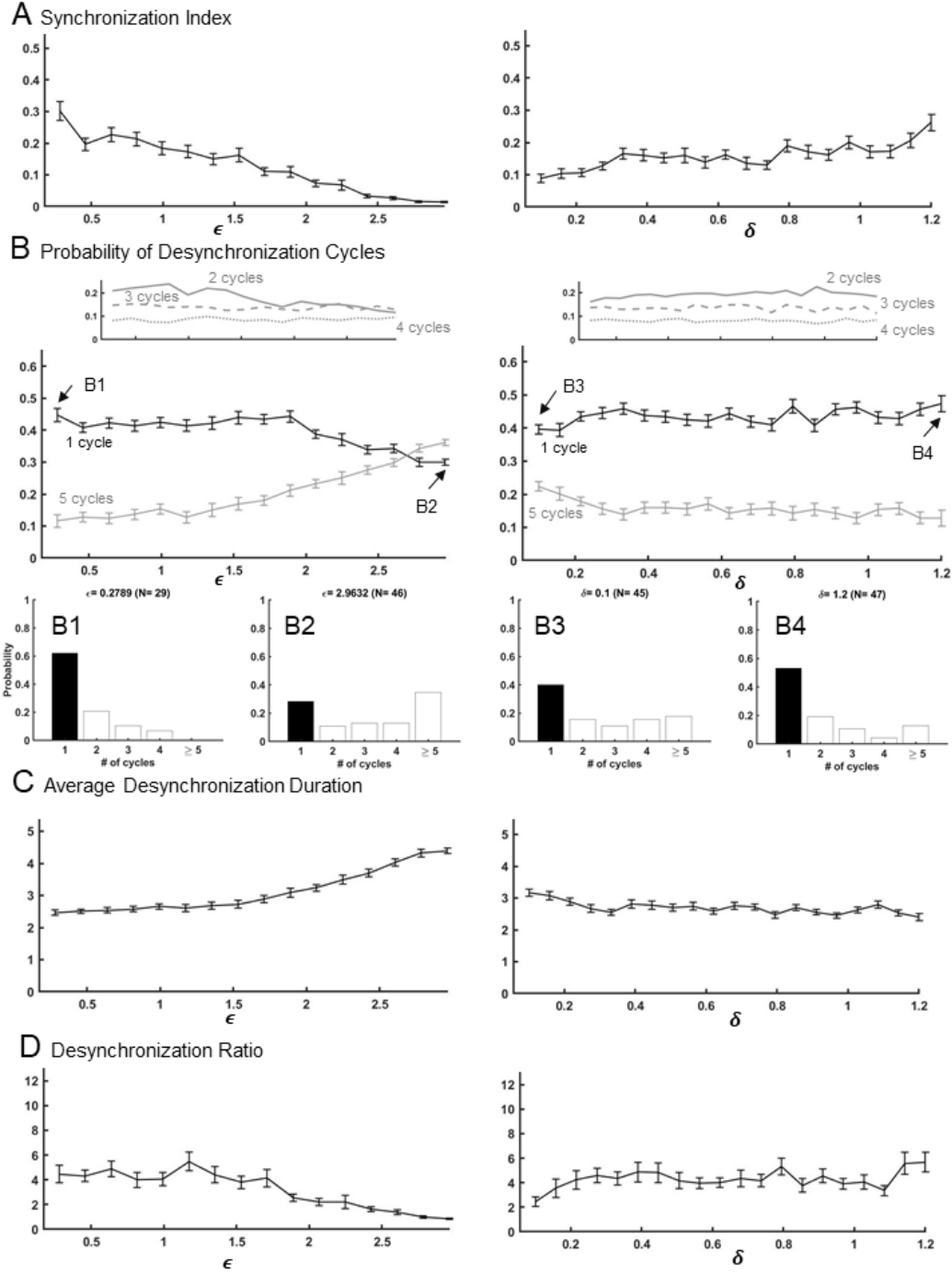

Figure 5: Effects of kinetic parameters on temporal patterning of synchronization. Variations of the peak value of activation time constant of potassium channel ($\varepsilon$) and the width of voltage-dependence of the activation time constant of potassium channel ($\delta$) are presented in the left and right columns, respectively. The results are shown as mean values across 50 simulations. The error bar is the standard error of the mean. (A) Synchronization strength index as a function of $\varepsilon$ and $\delta$. (B) Relative frequency of the modal value of the distribution of desynchronization durations as a function of $\varepsilon$ and $\delta$. The histograms (B1, B2, B3, B4) show examples of desynchronization distributions for the cases with longer and shorter desynchronizations. The black bin is the modal bin. The arrows indicate the parameter values of the examples. (C) Average desynchronization duration as a function of $\varepsilon$ and $\delta$. (D) Desynchronization ratio as a function of $\varepsilon$ and $\delta$.

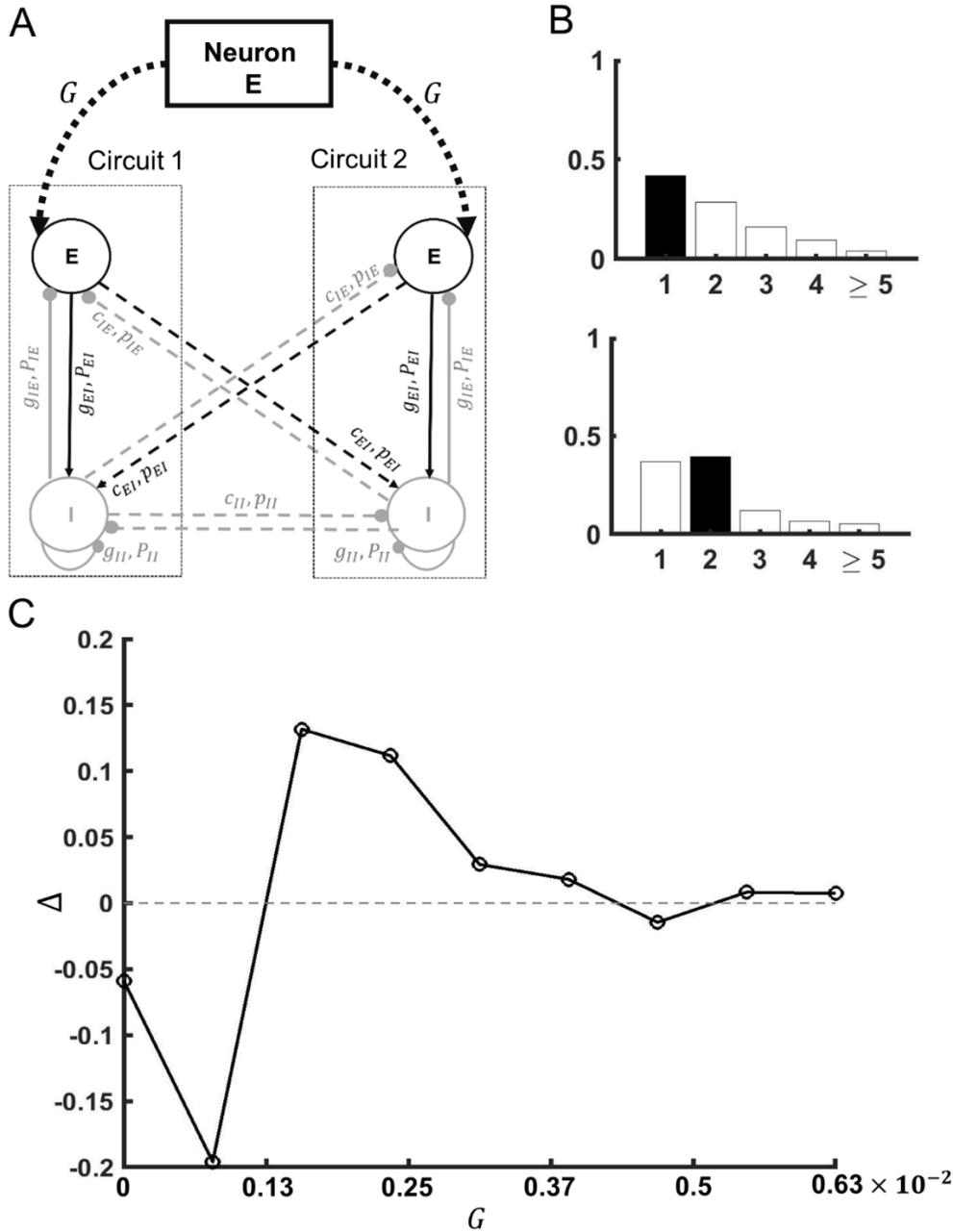

Figure 6: Different temporal patterning of synchronized dynamics yields different sensitivity to input signal. (A) Schematic of the network model receiving common synaptic input. The input is generated by one neuron and is sent to all excitatory neurons; the circuitry of the network receiving the input is the same as in Figure 1A. Here, $G$ represents the connection strength to one E neuron. (B) Histograms of desynchronization durations for "mode 1 system" (top) and "mode 2 system" (bottom). (C) The normalized synchrony difference $\Delta$ is plotted as a function of synaptic input strength $G$. Note that $G$ is the connection strength to one E neuron, and there are 40 E neurons in each circuit. Thus, the total connection strength is varied from 0 to 0.00625*40=0.25.

20